\documentclass[journal]{IEEEtran}
\ifCLASSINFOpdf
\else
\fi

\usepackage{multirow}
\usepackage{graphicx}
\usepackage{hyperref}
\usepackage{array, booktabs, longtable}	%for long long table
\usepackage{lineno,hyperref}
\begin{document}
	%
	% paper title
	% Titles are generally capitalized except for words such as a, an, and, as,
	% at, but, by, for, in, nor, of, on, or, the, to and up, which are usually
	% not capitalized unless they are the first or last word of the title.
	% Linebreaks \\ can be used within to get better formatting as desired.
	% Do not put math or special symbols in the title.
	\title{Skin Lesion Segmentation in Dermoscopic Images with Ensemble Deep Learning Methods}
	%
	%
	% author names and IEEE memberships
	% note positions of commas and nonbreaking spaces ( ~ ) LaTeX will not break
	% a structure at a ~ so this keeps an author's name from being broken across
	% two lines.
	% use \thanks{} to gain access to the first footnote area
	% a separate \thanks must be used for each paragraph as LaTeX2e's \thanks
	% was not built to handle multiple paragraphs
	%
	
	\author{Manu Goyal, Amanda Oakley, Priyanka Bansal, Darren Dancey,
		and~Moi Hoon Yap,~\IEEEmembership{Senior Member,~IEEE}% <-this % stops a space
		\IEEEcompsocitemizethanks{\IEEEcompsocthanksitem M. Goyal, P. Bansal, D. Dancey and M.H. Yap are with the Centre for Advanced Computational Sciences, Department of Computing and Mathematics,
			Manchester Metropolitan University, John Dalton Building, M1 5GD, Manchester, UK.\protect\\
			% note need leading \protect in front of \\ to get a newline within \thanks as
			% \\ is fragile and will error, could use \hfil\break instead.
			E-mail: M.Yap@mmu.ac.uk}
		\thanks{A. Oakley is with Waikato Clinical School, University of Auckland, Hamilton, New Zealand.}
		
		%	\IEEEcompsocthanksitem Fatima Mohamed Osman is with the Department of Computer Science, Sudan University of Science and Technology.}% <-this % stops an unwanted space
		\thanks{Manuscript received xxxxx xx, 2019; revised xxx xx, 2019.}}
	% note the % following the last \IEEEmembership and also \thanks - 
	% these prevent an unwanted space from occurring between the last author name
	% and the end of the author line. i.e., if you had this:
	% 
	% \author{....lastname \thanks{...} \thanks{...} }
	%                     ^------------^------------^----Do not want these spaces!
	%
	% a space would be appended to the last name and could cause every name on that
	% line to be shifted left slightly. This is one of those "LaTeX things". For
	% instance, "\textbf{A} \textbf{B}" will typeset as "A B" not "AB". To get
	% "AB" then you have to do: "\textbf{A}\textbf{B}"
	% \thanks is no different in this regard, so shield the last } of each \thanks
	% that ends a line with a % and do not let a space in before the next \thanks.
	% Spaces after \IEEEmembership other than the last one are OK (and needed) as
	% you are supposed to have spaces between the names. For what it is worth,
	% this is a minor point as most people would not even notice if the said evil
	% space somehow managed to creep in.

	% The paper headers
	\markboth{Journal of \LaTeX\ Class Files,~Vol.~XX, No.~X, XXXX~2017}%
	{Shell \MakeLowercase{\textit{et al.}}: Bare Demo of IEEEtran.cls for IEEE Journals}
	% The only time the second header will appear is for the odd numbered pages
	% after the title page when using the twoside option.
	% 
	% *** Note that you probably will NOT want to include the author's ***
	% *** name in the headers of peer review papers.                   ***
	% You can use \ifCLASSOPTIONpeerreview for conditional compilation here if
	% you desire.

	% If you want to put a publisher's ID mark on the page you can do it like
	% this:
	%\IEEEpubid{0000--0000/00\$00.00~\copyright~2015 IEEE}
	% Remember, if you use this you must call \IEEEpubidadjcol in the second
	% column for its text to clear the IEEEpubid mark.

	% use for special paper notices
	%\IEEEspecialpapernotice{(Invited Paper)}

	% make the title area
	\maketitle
	
	% As a general rule, do not put math, special symbols or citations
	% in the abstract or keywords.
	\begin{abstract}
		Early detection of skin cancer, particularly melanoma, is crucial to enable advanced treatment. Due to the rapid growth in the numbers of skin cancers, there is a growing need of computerized analysis for skin lesions. The state-of-the-art public available datasets for skin lesions are often accompanied with very limited amount of segmentation ground truth labeling as it is laborious and expensive. The lesion boundary segmentation is vital to locate the lesion accurately in dermoscopic images and lesion diagnosis of different skin lesion types. In this work, we propose the use of fully automated deep learning ensemble methods for accurate lesion boundary segmentation in dermoscopic images. We trained the Mask-RCNN and DeepLabv3+ methods on ISIC-2017 segmentation training set and evaluate the performance of the ensemble networks on ISIC-2017 testing set. Our results showed that the best proposed ensemble method segmented the skin lesions with Jaccard index of 79.58\%  for the ISIC-2017 testing set. The proposed ensemble method outperformed FrCN, FCN, U-Net, and SegNet in Jaccard Index by 2.48\%, 7.42\%, 17.95\%, and 9.96\% respectively. Furthermore, the proposed ensemble method achieved an accuracy of 95.6\% for some representative clinically benign cases, 90.78\% for the melanoma cases, and 91.29\% for the seborrhoeic keratosis cases on ISIC-2017 testing set, exhibiting better performance than FrCN, FCN, U-Net, and SegNet. 
	\end{abstract}
	
	% Note that keywords are not normally used for peerreview papers.

	% For peer review papers, you can put extra information on the cover
	% page as needed:
	% \ifCLASSOPTIONpeerreview
	% \begin{center} \bfseries EDICS Category: 3-BBND \end{center}
	% \fi
	%
	% For peerreview papers, this IEEEtran command inserts a page break and
	% creates the second title. It will be ignored for other modes.
	\IEEEpeerreviewmaketitle

	\section{Introduction and Background}
	% The very first letter is a 2 line initial drop letter followed
	% by the rest of the first word in caps.
	% 
	% form to use if the first word consists of a single letter:
	% \IEEEPARstart{A}{demo} file is ....
	% 
	% form to use if you need the single drop letter followed by
	% normal text (unknown if ever used by the IEEE):
	% \IEEEPARstart{A}{}demo file is ....
	% 
	% Some journals put the first two words in caps:
	% \IEEEPARstart{T}{his demo} file is ....
	% 
	% Here we have the typical use of a "T" for an initial drop letter
	% and "HIS" in caps to complete the first word.
	\IEEEPARstart{C}{ancers} of the skin are the most common cancers among all other cancers \cite{pathan2018techniques}. The most common malignant skin lesions are melanoma, squamous cell carcinoma and basal cell carcinoma. It is estimated that in 2019, 96,480 new cases will be diagnosed with melanoma and more than 7,000 people will die from the disease in the United States \cite{Seer2017} \cite{melanomafoundation2017}. Early detection of melanoma can save lives.

It can be difficult to differentiate benign lesions from skin cancers. Skin cancer specialists examine their patients' skin lesions using visual inspection aided by hand-held dermoscopy, and they may capture digital close-up (macroscopic) and dermoscopic (microscopic) images. Dermoscopy is a means to examine the skin using a bright light, magnification, and employs either polarisation or immersion fluid to reduce surface reflection \cite{pellacani2002comparison}.   In common use for the last 20 years, dermoscopy has improved the diagnosis rate over visual inspection alone \cite{mayer1997systematic}. The ABCD criteria were devised in 1987 to help non-dermatologists screen skin lesions to differentiate common benign melanocytic naevi (naevi) from melanoma \cite{abbasi2004early}. It does not employ dermoscopy. Fig. 1 illustrates the ABCD rules for skin lesion diagnosis, where:
	
	\begin{figure}
		\centering
		\includegraphics[scale=0.45]{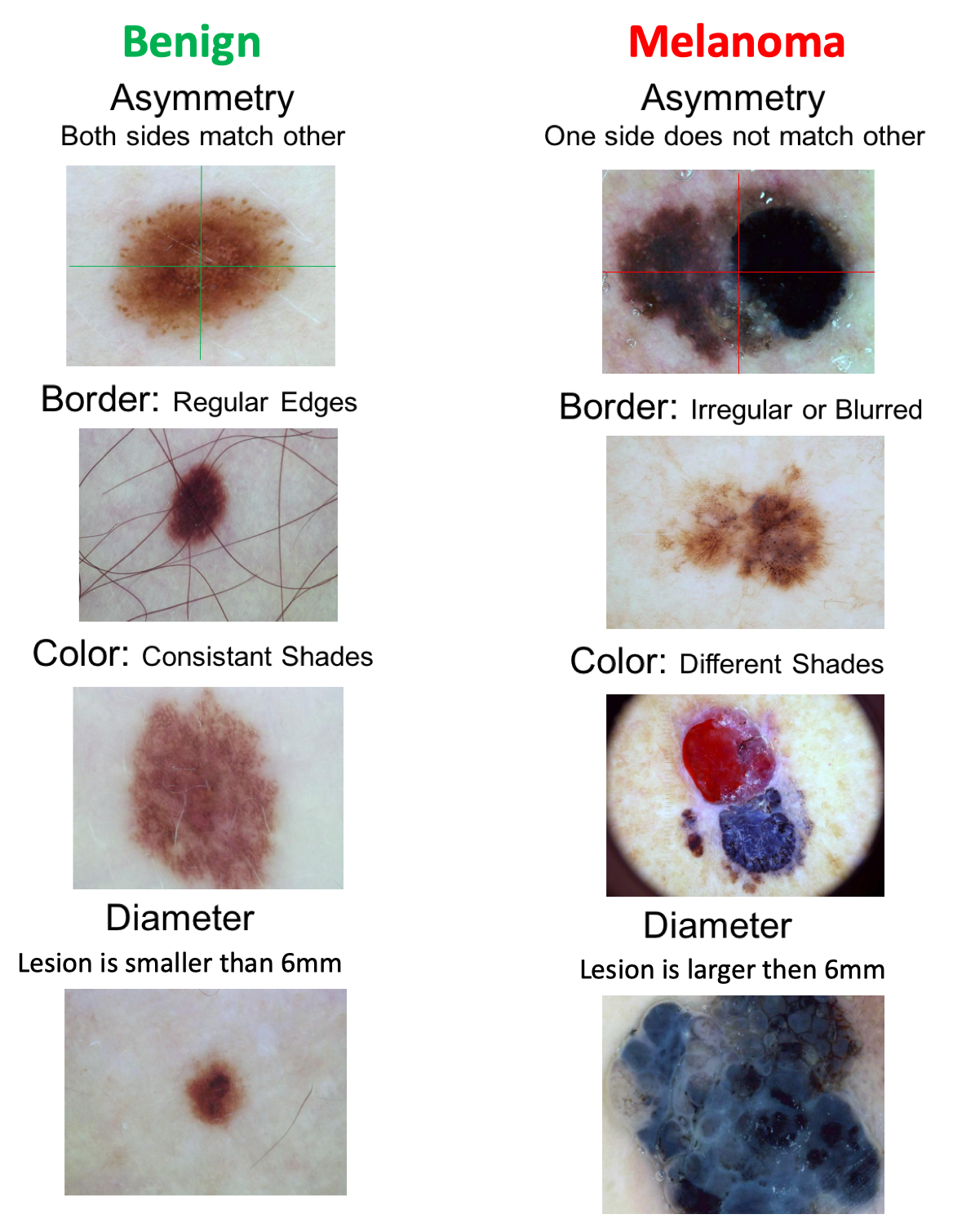}
		\caption{Lesion Diagnosis by Dermatologists. ABCD Criteria for lesion diagnosis focuses on finding the certain properties of lesions}
		\label{fig:ABCD}
	\end{figure}
	\begin{enumerate}
		\item A:  Asymmetry  property checks whether two halves of the skin lesion match or not in terms of colour, shape, edges. The skin lesions are divided into two halves based on long axis and short axis as shown in the Fig. 1. In the case of melanoma, it is likely to have an asymmetrical appearance.
		
		\item B: Border  property. It defines whether the edges of skin lesion are smooth,
well-defined or otherwise. In the case of melanoma, edges are likely to be uneven, blurry and jagged.

		\item C: Colour  property. The colour in a melanoma, varies from one  area to the another, and it often has different shades of tan, brown, red, and black.
		
		\item D: Diameter  property. It measures the approximate diameter of the skin lesion. The diameter of a melanoma is generally greater than 6mm (the size of pencil eraser).
	\end{enumerate}
	
	End-to-end computerized solutions that can produce accurate segmentation of skin lesions irrespective of types of skin lesions are highly desirable to mirror the ABCD Rule. For segmentation of medical imaging, \textit{dice}, \textit{specificity} and \textit{sensitivity} are deemed as important performance measures for methods. Hence, computerized methods need to achieve high scores in these performance metrics. 
	
	The majority of the state-of-the-art computer-aided diagnosis using dermoscopy images is multi-stage, and includes image pre-processing, image segmentation, features extraction and classification \cite{korotkov2012computerized, pathan2018techniques}. 	Using hand-crafted feature descriptors, benign naevi tend to have small dimensions and a roundish shape, as illustrated in Fig. 1. (but some naevi are large and unusual shapes). Other feature descriptors used in previous works include asymmetry features, colour features and texture features. Pattern analysis is widely used to describe the dermoscopic appearance of skin lesions, for example the melanocytic algorithm elaborated by Argenziano et al. \cite{argenziano2000interactive}. Various computer algorithms have been devised to classify lesion types using features descriptors and pattern analysis based on image processing and conventional machine learning approaches. Two reviews by Korotkov et al. \cite{korotkov2012computerized} and Pathan et al. \cite{pathan2018techniques} reported that the majority of these used hand-crafted features to classify or segment the lesions. Korotkov et al. \cite{korotkov2012computerized} concluded that there
 is a large discrepancy in previous research and the computer-aided diagnosis (CAD) systems were not ready for implementation. The other issue was the lack of a benchmark dataset, which makes it harder to assess the algorithms. Pathan et al. \cite{pathan2018techniques} concluded that the CAD systems worked in experimental settings but required rigorous validation in real-world clinical settings.

	With the rapid growth of deep learning approaches, many researchers \cite{yuan2017automatic, yu2017automated, bi2017dermoscopic} have proposed using Deep Convolutional Neural Networks (CNN) for melanoma detection and segmentation. We designed a fully automatic CNN-based ensemble method for accurate lesion boundary segmentation and trained it on the ISIC-2017 dermoscopic training set. Then, we tested the robustness of the ISIC-2017 trained algorithms on another publicly available dataset, the PH2 dataset.
	
	\section{Deep Learning for Skin Lesion Segmentation}
	Deep learning has gained popularity in medical imaging research including Magnetic Resonance Imaging (MRI) on brain \cite{zhang2015deep}, breast ultrasound cancer detection \cite{yap2017automated} and diabetic foot ulcer classification and segmentation \cite{goyal2017dfunet, goyal2017fully}. U-Net is a popular deep learning approach in biomedical imaging research, proposed by Ronneberger et al. \cite{ronneberger2015u}. U-Net enables the use of data augmentation, including the use of non-rigid deformations, to make full use of the available annotated sample images to train the model. These aspects suggest that the U-Net could potentially provide satisfactory results with the limited size of the biomedical datasets currently available. An up-to-date review of conventional machine learning methods is presented in \cite{pathan2018techniques}. This section reviews the state-of-the-art deep learning approaches for segmentation for skin lesions. 
	
	Researchers have made significant contributions proposing various deep learning frameworks for the detection of skin lesions. Yu et al. \cite{yu2017automated} proposed very deep residual networks of more than 50 layers for two-stage framework of skin lesions segmentation followed by classification. They claimed that the deeper networks produce richer and more discriminative features for recognition. By validating their methods on ISBI 2016 \textit{Skin Lesion Analysis Towards Melanoma Detection Challenge} dataset \cite{gutman2016skin}, they reported that their method ranked first in classification when compared to 16-layer VGG-16, 22-layer GoogleNet and other 25 teams in the competition. However, in segmentation stage, they ranked second in segmentation among the 28 teams. Although the work showed promising results, but the two-stage framework and very deep networks were computationally expensive.
	
			\begin{figure}[!t]
		\centering
		\begin{tabular}{ccc}
			\includegraphics[width=4cm,height=3cm]{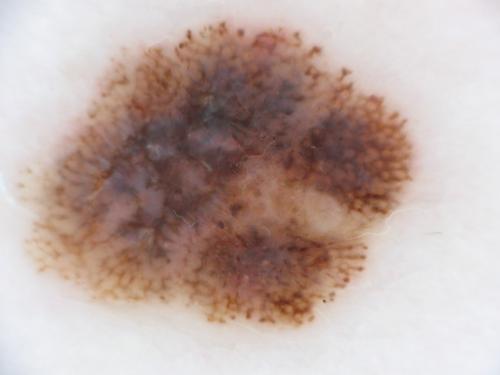} &
			\includegraphics[width=4cm,height=3cm]{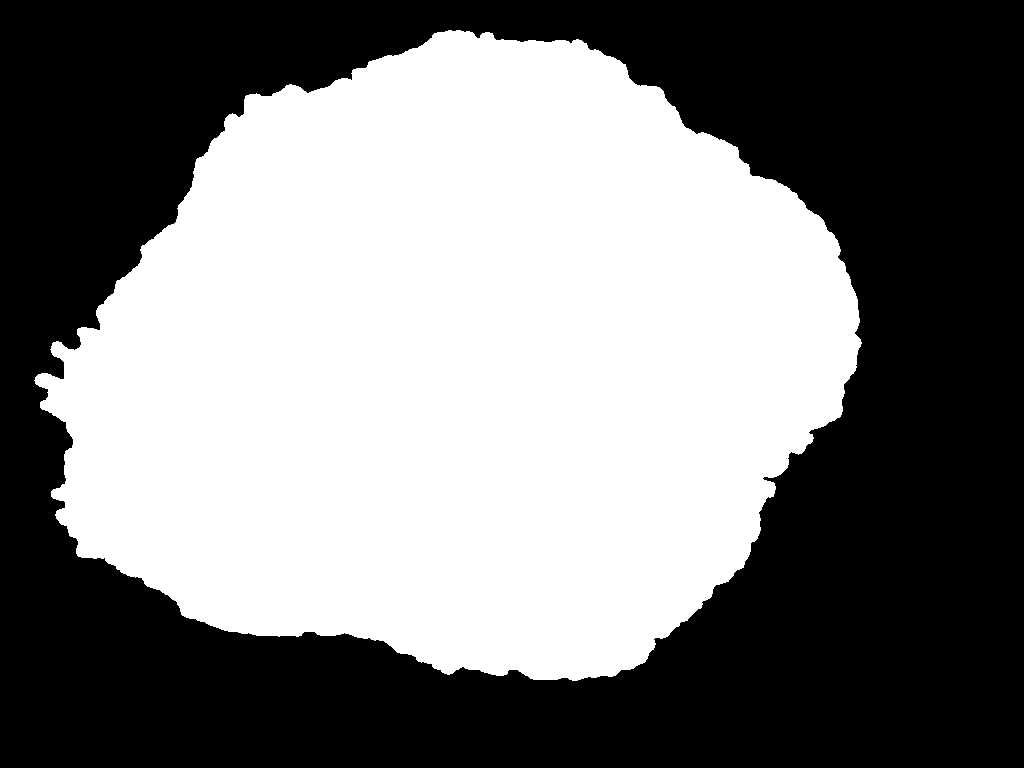}    
			\\
			\includegraphics[width=4cm,height=3cm]{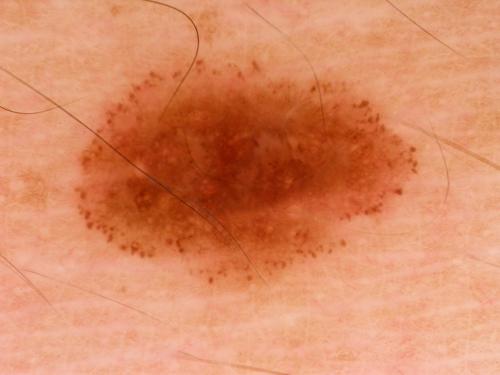} &
			\includegraphics[width=4cm,height=3cm]{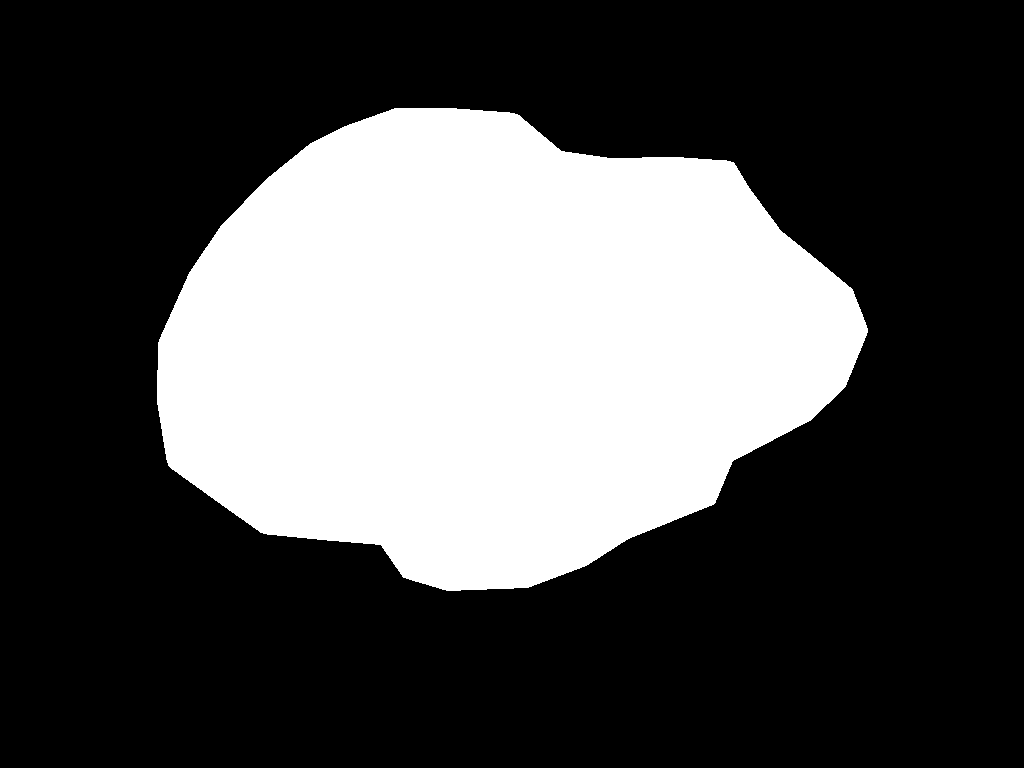}
			\\
		\end{tabular}     
		
		\caption{Examples from skin lesion dataset. (Left) Original Images; and (Right) Ground truth in binary masks.}
		\label{fig:masks}
	\end{figure}
	
	\begin{figure*}
		\centering
		\includegraphics[scale=0.5]{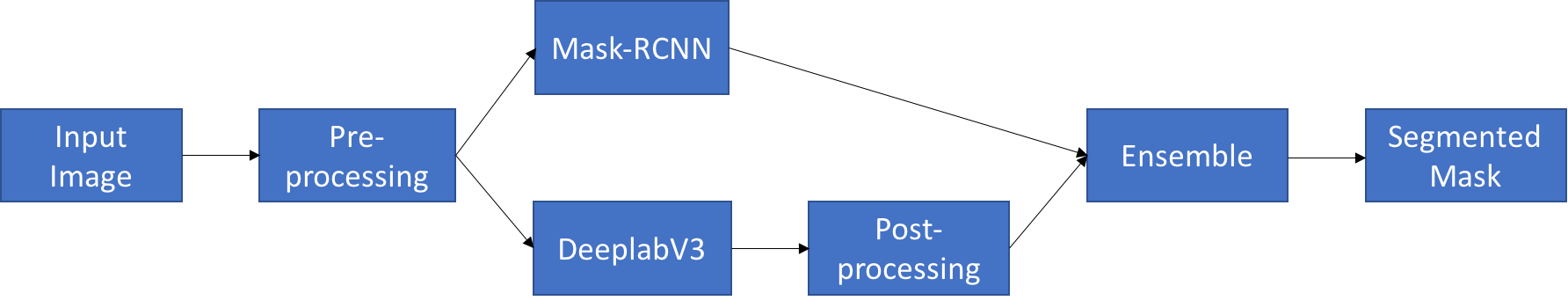}
		\caption{Complete flow of our proposed ensemble methods for automated skin lesion segmentation.}
		\label{fig:Arch}
	\end{figure*}

	Bi et al. \cite{bi2017dermoscopic} proposed a multi-stage fully convolutional networks (FCNs)for skin lesions segmentation. The multi-stage involved localised coarse appearance learning in the early stage and detailed boundaries characteristics learning in the later stage. Further, they implemented a parallel integration approach to enable fusion of the result that they claimed that this has enhanced the detection. Their method outperformed others in PH2 dataset \cite{mendoncca2013ph} of 90.66\% but achieved marginal improvement if compared to Team ExB in ISIB 2016 competition with 91.18\%. 
	
	Yuan et al. \cite{yuan2017automatic} proposed an end-to-end fully automatic method for skin lesions segmentation by leveraging 19-layer DCNN. They introduced a loss function using Jaccard Distance as the measurement. They compared the results using different parameters such as input size, optimisation methods, augmented strategies, and loss function. To fine tune the hyper-parameters, 5-fold cross-validation with ISBI training dataset was used to determine the best performer. Similar to Bi et al. \cite{bi2017dermoscopic}, they evaluated their results on ISBI 2016 and PH2 dataset. The results were outperformed by the state-of-the-art methods but they suggested that the method achieved poor results in some challenging cases including images with low contrast.
	
	Goyal et al. \cite{goyal2017multi} proposed fully convolutional methods for multi-class segmentation on ISBI challenge dataset 2017. This was a very first attempt to perform multi-class segmentation to distinguish melanocytic naevus, melanoma and seborrhoeic keratoses rather than single class of skin lesion.  
	
	The research showed that deep learning achieved promising results for skin lesions segmentation and classification. However, these methods did not make their codes available and not validated on the ISIC-2017 dataset, which has 2000 images compared to 900 in ISIC-2016 dataset.	
		
	\section{Methodology}
	This section discusses the publicly available skin lesion datasets, the preparation of the ground truth, and the performance measures to validate our results. 
		
	\subsection{Skin Lesion Datasets}
	For this work, we used two publicly available datasets for skin lesions, which are ISIC-2017 Challenge (Henceforth ISIC-2017) \cite{codella2018skin} and PH2 dataset \cite{mendoncca2013ph}. ISIC-2017 is a subset of ISIC Archive dataset \cite{codella2017skin}. In segmentation category, it consists of 2750 images with 2000 images in training set, 150 in validation set and 600 in the testing set.  Even though ISIC Challenge 2018 \cite{codella2018skin} was conducted last year, they did not share the ground truth of their testing set. Therefore, our work was based on the ISIC-2017. PH2 has 200 images in which 160 images are naevus (atypical naevus and common naevus), and 40 images are of melanoma. The ground truth for both datasets is in a form of binary mask, as shown in the Fig. \ref{fig:masks}. We used the ISIC-2017 training set to build the prediction models and test the performance on ISIC-2017 testing set and PH2 dataset. To improve the performance and reduce the computational cost, we resized all the images to 500 $\times$ 375. 
		
	\subsection{Ensemble Methods for Lesion Boundary Segmentation}
	We designed this end-to-end ensemble segmentation method to combine Mask-RCNN and DeeplabV3+ with pre-processing and post-processing method to produce accurate lesion segmentation as shown in the Fig. \ref{fig:Arch}. This section describes each stage of our proposed ensemble method.

	\subsubsection{Pre-Processing}
\begin{figure}[!t]
		\centering
		\begin{tabular}{ccc}
			\includegraphics[width=4cm,height=3cm]{ISIC_0000020.jpg} &
			\includegraphics[width=4cm,height=3cm]{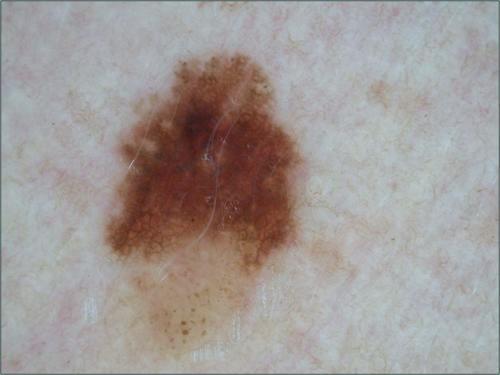}    
			\\
			\includegraphics[width=4cm,height=3cm]{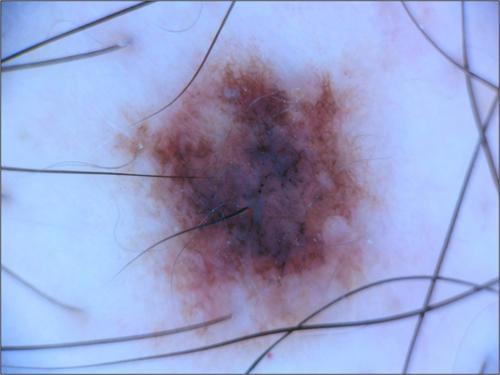} &
			\includegraphics[width=4cm,height=3cm]{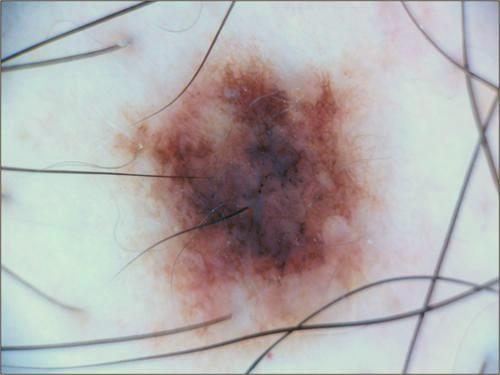}
			\\
			(a) Original images& (b) Pre-processed images \\ 
		\end{tabular}     
		
		\caption{Examples of pre-processing stage by using Shades of Gray algorithm. (a) Original images with different background colours; and (b) Pre-processed images with more consistent background colours.}
		\label{fig:preprocess}
	\end{figure}

	The ISIC Challenge dataset comprised of dermoscopic skin lesion images taken by different dermatoscope and camera devices all over the world. Hence, it is important to perform pre-processing for colour normalization and illumination with colour constancy algorithm \cite{ng2019effect}. We processed the datasets with Shades of Gray algorithm \cite{finlayson2004shades} as shown in Fig. \ref{fig:preprocess}. 
	
	\subsubsection{DeepLabv3+}
\begin{figure}
		\centering
		\includegraphics[scale=0.3]{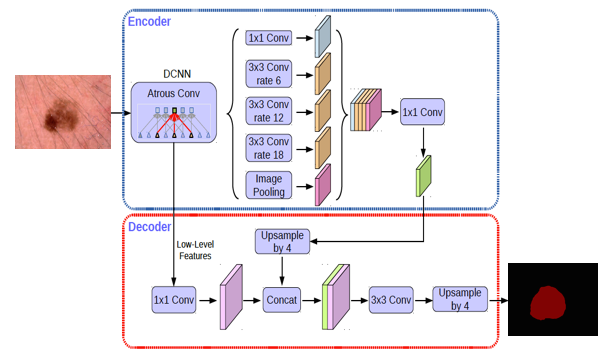}
		\caption{Architecture of DeepLabv3+ on skin lesion segmentation \cite{chen2018deeplab}.}
		\label{fig:deepArch}
	\end{figure}

	We trained DeepLabv3+ with default setting on the skin lesion datasets, which is one of the best performing semantic segmentation networks \cite{chen2018deeplab}. It assigns semantic label lesion to every pixel in a dermoscopic image. DeepLabv3+ is an encoder-decoder network which makes the use of CNN called Xception-65 with atrous convolution layers to get the coarse score map and then, conditional random field is used to produce final output as shown in Fig. \ref{fig:deepArch}.
	
	\subsubsection{Mask-RCNN}
	We fine-tuned Mask-RCNN with ResNet-InceptionV2 (henceforth Mask-RCNN) for single class as skin lesion for this experiment \cite{he2017mask}. In default setting, in some cases, Mask-RCNN generates more than one output. We modified the final layer of Mask-RCNN architecture to produce only single output mask of highest confidence per image. 
	
	\subsubsection{Post-processing}
\begin{figure}
		\centering
		\begin{tabular}{ccc}
			\includegraphics[width=4cm,height=3cm]{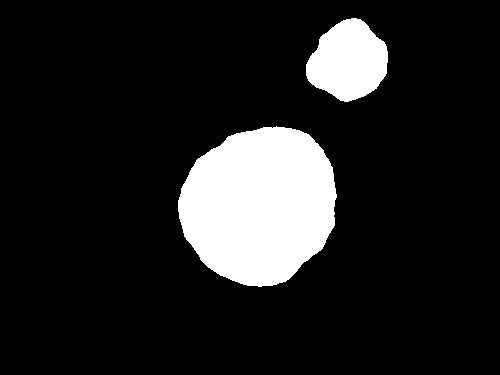} &
			\includegraphics[width=4cm,height=3cm]{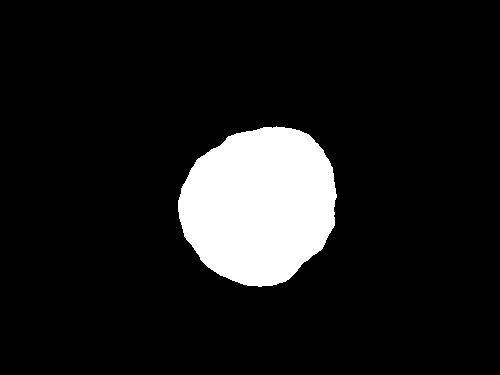}    
			\\
			\includegraphics[width=4cm,height=3cm]{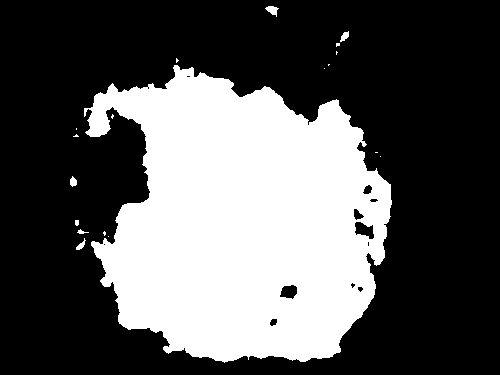} &
			\includegraphics[width=4cm,height=3cm]{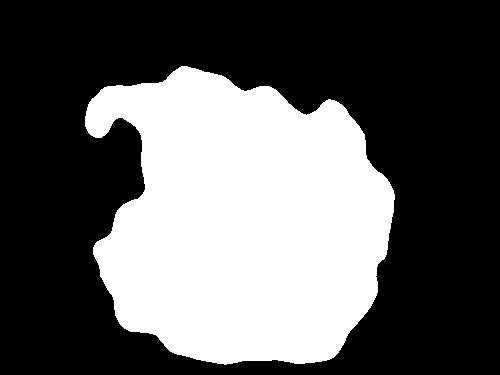}
			\\
			(a) Results from CNN & (b) Post-processed masks \\
		\end{tabular}     
		
		\caption{Examples of post-processing stage by using image processing methods: (a) Results from CNN segmentation with artefacts and holes within the lesions; and (b) Post-processed result after morphology operations.}
		\label{fig:postprocess}
	\end{figure}

	We used basic image processing methods, i.e. morphological operations to fill the region and remove unnecessary artefacts of the results as illustrated in the Fig. \ref{fig:postprocess}. These issues were only countered by DeepLabv3+ as in the case of Mask-RCNN, we have not had these issues. Hence, post-processing is only used for the semantic segmentation methods like FCN and DeepLabv3+. 
	
	\subsubsection{Ensemble Methods}
\begin{figure}[!t]
		\centering
		\begin{tabular}{cc}
			\includegraphics[width=8cm,height=6.25cm]{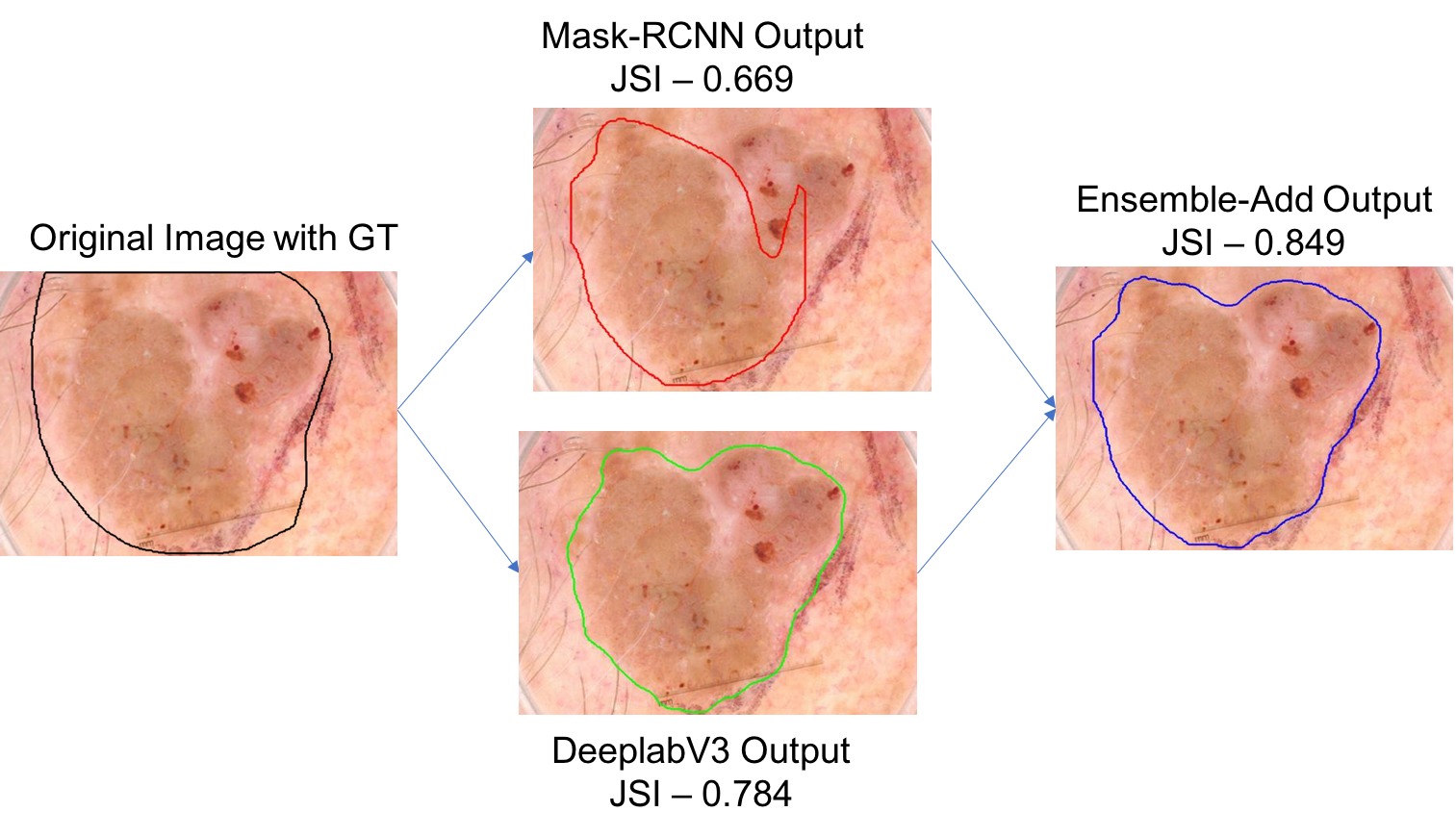}  \\
			(a) Ensemble - ADD \\
			\includegraphics[width=8cm,height=6.25cm]{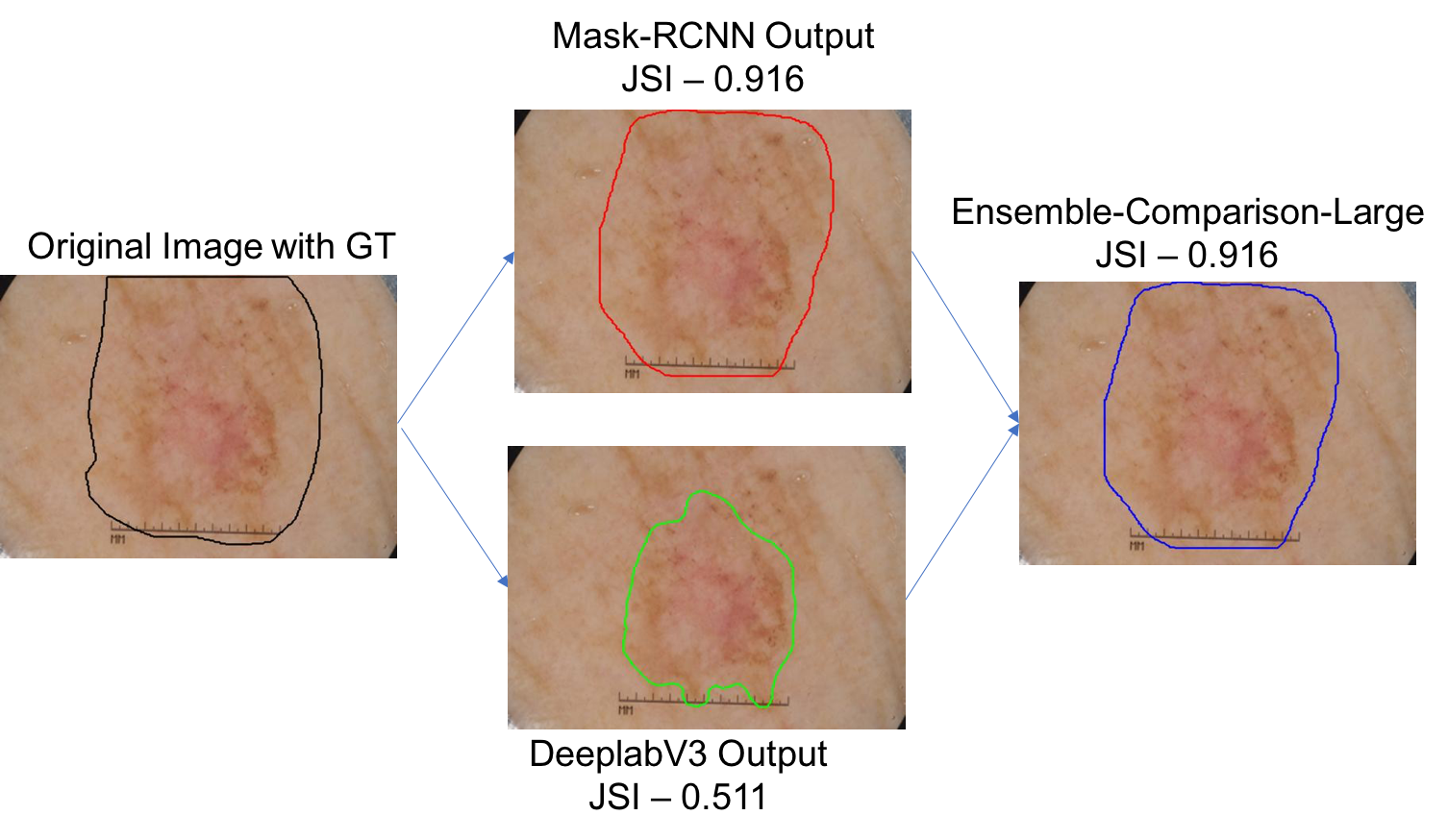}    \\
			(b) Ensemble - Comparison-Large\\
			\includegraphics[width=8cm,height=6.25cm]{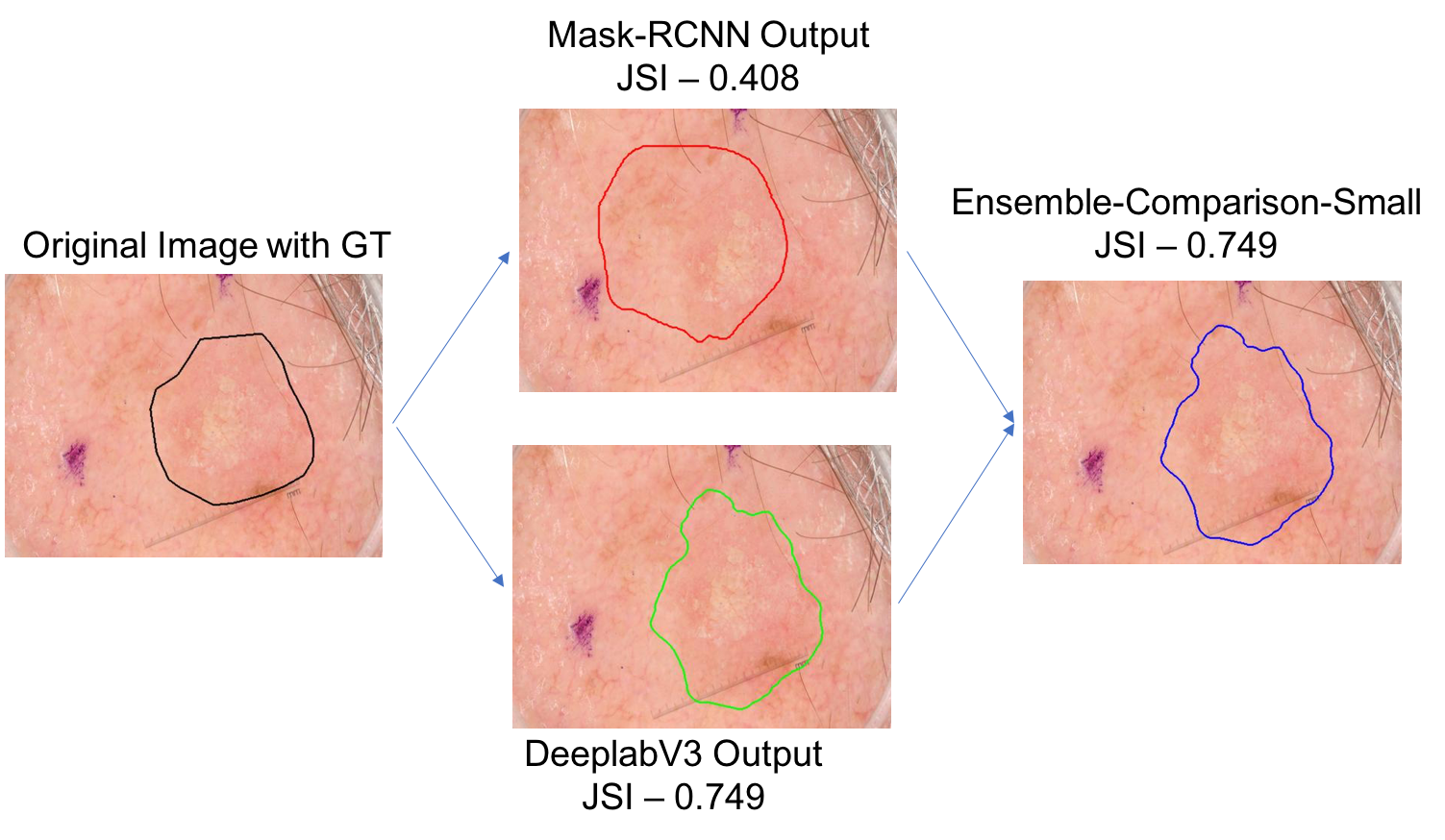}    \\
			(c) Ensemble - Comparison-Small\\
		\end{tabular}     
		
		\caption{Illustration of ensemble methods: (a) Ensemble-ADD (b) Ensemble-Comparison (Large) (c) Ensemble-Comparison (Small)}
		\label{fig:ensemble}
	\end{figure}

	We used two types of ensemble methods called Ensemble-ADD and Ensemble-Comparison. First of all, if there is no prediction from DeepLabv3+, ensemble method picks up the prediction of Mask-RCNN and vice versa. Then, Ensemble-ADD combines the results of both Mask-RCNN and DeepLabv3+ to produce final segmentation mask. Ensemble-Comparison-Large picks the larger segmented area by comparing the number of pixels in output of both methods. In contrary, Ensemble-Comparison-Small picks the smaller area from the output. The ensemble methods are illustrated by Fig. \ref{fig:ensemble} where (a) shows Ensemble-ADD; (b) shows Ensemble-Comparison-Large; and (c) represents Ensemble-Comparison-Small.
	
			\begin{table*}[]
		\centering
		\addtolength{\tabcolsep}{2pt}
		\renewcommand{\arraystretch}{1.5}
		\caption{Performance evaluation of our proposed methods and state-of-the-art algorithms on ISIC Skin Lesion Segmentation Challenge 2017}
		\label{my-label2}
		\scalebox{0.92}{
			\begin{tabular}{cccccc}
				\hline
				Method     & Accuracy & Dice &Jaccard Index &Sensitivity&Specificity  \\ \hline \hline
				First: Yading Yuan (CDNN Model)   &0.934 &0.849    & 0.765 & 0.825 & 0.975\\ 
				Second: Matt Berseth (U-Net) &0.932 &0.847    & 0.762 & 0.820 & 0.978\\  
				U-Net \cite{ronneberger2015u} &0.901 &0.763    & 0.616 & 0.672 & 0.972\\ 
				SegNet \cite{badrinarayanan2015segnet} &0.918 &0.821    & 0.696 & 0.801 & 0.954\\
				FrCN \cite{al2018skin} &0.940 &0.870    & 0.771 & 0.854 & 0.967\\ 
				Ensemble-S (Proposed Method)   &0.933 &0.844    & 0.760 & 0.806 & \textbf{0.979} \\ 
				Ensemble-L (Proposed Method) &0.939 &0.866    & 0.788 & 0.887 & 0.955 \\ 
				Ensemble-A (Proposed Method) &\textbf{0.941} & \textbf{0.871}    & \textbf{0.793} & \textbf{0.899} & 0.950 \\ \hline        
		\end{tabular}}
	\end{table*}  	

	\begin{figure*}
		\centering
		\includegraphics[scale=0.75]{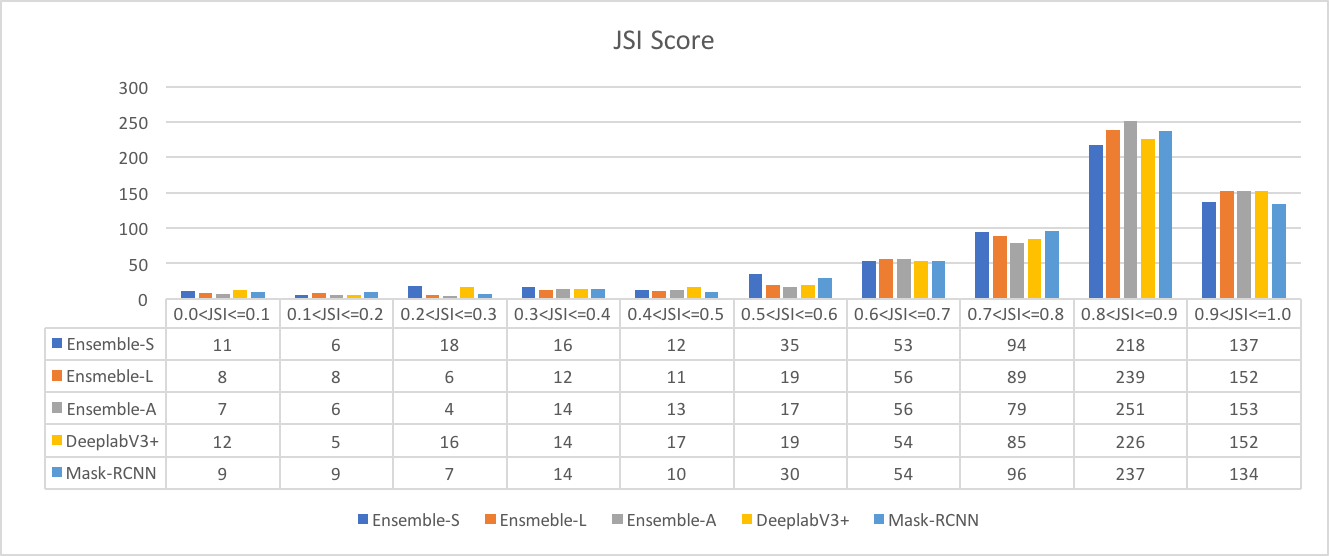}
		\caption{Comparison of JSI scores of our proposed methods for skin lesion segmentation of ISIC-2017 testing set.}
		\label{fig:JSI_Comp}
	\end{figure*}
	
	\subsection{Performance Metrics}
	% Description about Dice, Sensitivity, Specifity, FROC
	We evaluated the performance of the segmentation algorithms by using \textit{Dice Similarity Coefficient (Dice)} \cite{zikic2014segmentation, pereira2016brain}. In addition, we report our findings in \textit{Jaccard Similarity Index (JSI)}, \textit{Sensitivity}, \textit{Specificity}, \textit{Accuracy} and \textit{Matthew Correlation Coefficient (MCC)} \cite{powers2011evaluation}. 
	
	\begin{equation}
	Sensitivity= \frac {TP}{TP+FN}
	\end{equation}
	
	\begin{equation}
	Specificity= \frac {TN}{FP+TN}
	\end{equation}
	
	\begin{equation}
	Accuracy= \frac {TP+TN}{TP+FP+TN+FN}
	\end{equation}
	
	\begin{equation}
	JSI= \frac {TP}{(TP + FP + FN)}
	\end{equation}
	
	\begin{equation}
	Dice= \frac {2*TP}{(2*TP + FP + FN)}
	\end{equation}
	
	\begin{small}
		\begin{equation}
		MCC= \frac {TP*TN - FP*FN}{\sqrt{(TP+FP)(TP+FN)(TN+FP)(TN+FN)}}
		\end{equation}
	\end{small}
	
	\textit{Sensitivity} is defined in eq (1), where \textit{TP} is \textit{True Positives} and \textit{FN} is \textit{False Negatives}. A high \textit{Sensitivity} (close to 1.0) indicates good performance in segmentation which implies all the lesions were segmented successfully. On the other hand, \textit{Specificity} (as in eq. (2)) indicates the proportion of \textit{True Negatives (TN)} of the non-lesions. A high \textit{Specificity} indicates the capability of a method in not segmenting the non-lesions. \textit{JSI} and \textit{Dice Similarity Index (Dice)} is a measure of how similar both prediction and ground truth are, by measuring of how many \textit{TP} found and penalising for the \textit{FP} that the method found, as in eq. (3). \textit{MCC} has a range of -1 (completely wrong binary classifier) to 1 (completely right binary classifier). This is a suitable measurement for the performance assessment of our segmentation algorithms based on binary classification (lesion versus non-lesions), as in eq. (4).
	
		\begin{table*}[]
		\centering
		\small\addtolength{\tabcolsep}{2pt}
		\renewcommand{\arraystretch}{1.5}
		\caption{Performance evaluation of our proposed methods and state-of-the-art segmentation architectures on ISIC 2017 testing set (SEN denotes \textit{Sensitivity},SPE is \textit{Specificity}, ACC is \textit{Accuracy}, and SK denotes Seborrhoeic Keratosis)}
		\label{tab:tradFeatsn}
		\scalebox{0.82}{
			\begin{tabular}{|c|c|c|c|c|c|c|c|c|c|c|c|c|c}
				\cline{1-13}
				\multirow{2}{*}{Method} & \multicolumn{3}{c|}{Naevus}                              & \multicolumn{3}{c|}{Melanoma}                       & \multicolumn{3}{c|}{SK}                       & \multicolumn{3}{c|}{Overall}                               &  \\ \cline{2-13}
				& SEN & SPE & ACC    & SEN    & SPE   & ACC      & SEN    & SPE   & ACC      & SEN & SPE & ACC     &  \\ \cline{1-13}
				FCN-AlexNet & 82.44  & 97.58    & 94.84 & 72.35     & 96.23      & 87.82   & 71.70     & 97.92     & 89.35   & 78.86  & 97.37    & 92.65  &  \\ \cline{1-13}
				FCN-32s & 83.67  & 96.69    & 94.59 & 74.36     & 96.32      & 88.94   & 75.80     & 96.41      & 89.45   & 80.67  & 96.72    & 92.72  &  \\ \cline{1-13}
				FCN-16s & 84.23  & 96.91    & 94.67 & 75.14     & 96.27      & 89.24   & 75.48    & 96.25      & 88.83   & 81.14  & 96.68    & 92.74  &  \\ \cline{1-13}
				FCN-8s & 83.91  & 97.22    & 94.55 & 78.37     & 95.96      & 89.63   & 69.85     & 96.57      & 87.40   & 80.72  & 96.87    & 92.52 &  \\ \cline{1-13}
				DeeplabV3+  & 88.54  & 97.21    & \textbf{95.67} & 77.71     & 96.37      & 89.65   & 74.59     & 98.55      & 90.06   & 84.34  & 97.25    & 93.66  &  \\ \cline{1-13}
				Mask-RCNN             & 87.25  & 96.38    & 95.32 & 78.63     & 95.63      & 89.31   & 82.41     & 94.88     & 90.85   & 84.84  & 96.01    & 93.48  &  \\ \cline{1-13}
				Ensemble-S & 84.74  & \textbf{97.98}    & 95.58 & 73.35     & \textbf{97.30}      & 88.40   & 71.80     & \textbf{98.58}      & 89.91   & 80.58  & \textbf{97.94}    & 93.33 &  \\ \cline{1-13}
				Ensemble--L & 90.93  & 95.74    & 95.51 & 83.40     & 95.00      & 90.61   & 85.81     & 94.74      & 91.34   & 88.70  & 95.45    & 93.93 &  \\ \cline{1-13}
				Ensemble-A  & \textbf{92.08}  & 95.37    & 95.59 & \textbf{84.62}     & 94.20      & \textbf{90.85}   & \textbf{87.48}     & 94.41      & \textbf{91.72}   & \textbf{89.93}  & 95.00    & \textbf{94.08}  &  \\ \cline{1-13}
		\end{tabular}}
	\end{table*}
	\begin{table*}[]
		\centering
		\small\addtolength{\tabcolsep}{2pt}
		\renewcommand{\arraystretch}{1.5}
		\caption{Performance evaluation of our proposed methods and state-of-the-art segmentation architectures on ISIC 2017 testing set (DIC denotes \textit{Dice Score},JSI is \textit{Jaccard Similarity Index}, MCC is \textit{Mathews Correlation Coefficient}, and SK denotes Seborrhoeic Keratosis)}
		\label{tab:tradFeats}
		\scalebox{0.82}{
			\begin{tabular}{|c|c|c|c|c|c|c|c|c|c|c|c|c|c}
				\cline{1-13}
				\multirow{2}{*}{Method} & \multicolumn{3}{c|}{Naevus}                              & \multicolumn{3}{c|}{Melanoma}                       & \multicolumn{3}{c|}{SK}                       & \multicolumn{3}{c|}{Overall}                               &  \\ \cline{2-13}
				& DIC & JSI & MCC    & DIC    & JSI   & MCC      & DIC    & JSI   & MCC      & DIC & JSI & MCC     &  \\ \cline{1-13}
				FCN-AlexNet  & 85.61  & 77.01    & 82.91 & 75.94     & 64.32      & 70.35   & 75.09     & 63.76     & 71.51   & 82.15  & 72.55    & 78.75  &  \\ \cline{1-13}
				FCN-32s & 85.08  & 76.39    & 82.29 & 78.39     & 67.23      & 72.70   & 76.18     & 64.78      & 72.10   & 82.44  & 72.86    & 78.89  &  \\ \cline{1-13}
				FCN-16s & 85.60  & 77.39    & 82.92 & 79.22     & 68.41      & 73.26   & 75.23     & 64.11      & 71.42   & 82.80  & 73.65    & 79.31  &  \\ \cline{1-13}
				FCN-8s  & 84.33  & 76.07    & 81.73 & 80.08     & 69.58      & 74.39   & 68.01     & 56.54      & 65.14   & 81.06  & 71.87    & 77.81 &  \\ \cline{1-13}
				DeeplabV3+  & 88.29  & 81.09    & 85.90 & 80.86     & 71.30      & 76.01   & 77.05     & 67.55      & 74.62   & 85.16  & 77.15    & 82.28  &  \\ \cline{1-13}
				Mask-RCNN            & 88.83  & 80.91    & 85.38 & 80.28     & 70.69      & 74.95   & 80.48     & 70.74     & 76.31   & 85.58  & 77.39    & 81.99  &  \\ \cline{1-13}
				Ensemble-S    & 87.93  & 80.46    & 85.58 & 78.45     & 68.42      & 73.61   & 76.88     & 66.62      & 74.05   & 84.42  & 76.03    & 81.51 &  \\ \cline{1-13}
				Ensemble-L  & 88.87  & 81.69    & 85.93 & 83.05     & 74.01      & 77.98   & 81.71     & 72.50      & 77.68   & 86.66  & 78.82    & 83.14 &  \\ \cline{1-13}
				Ensemble-A   & \textbf{89.28}  & \textbf{82.11}    & \textbf{86.33} & \textbf{83.54}     & \textbf{74.53}      & \textbf{78.08}   & \textbf{82.53}     & \textbf{73.45}      & \textbf{78.61}   & \textbf{87.14}  & \textbf{79.34}    & \textbf{83.57}  &  \\ \cline{1-13}
				
		\end{tabular}}
	\end{table*}

	\section{Experiment and Results} 
	This section presents the performance of our proposed methods and various state-of-the-art segmentation methods on ISIC-2017 testing set (600 images) and PH2 dataset (200 images) \cite{codella2017skin, mendoncca2013ph}. 
	
	We train all the networks on a GPU machine with the following specification: (1) Hardware: CPU - Intel i7-6700 @ 4.00Ghz, GPU - NVIDIA TITAN X 12Gb, RAM - 32GB DDR5 (2) Software: Tensor-flow. 	
	
	\subsection{Comparison with ISIC Challenge 2017}
	Table \ref{my-label2} summarizes the performance of our proposed methods when compared to the best method in the ISIC-2017 segmentation challenge and other segmentation algorithms presented in \cite{al2018skin}. We compared our results using default competition performance metrics. Our proposed methods achieved highest scores in the default performance measures in this challenge when compared to the other algorithms. Our proposed method Ensemble-Add achieved \textit{Jaccard Similarity Index} of 79.34\% for ISIC testing set 2017 which outperformed U-Net, SegNet, and FrCN. Ensemble-S outperformed other algorithms in terms of textit{Specificity} with the score of 97.94\% where as  Ensemble-A received highest score in \textit{Sensitivity} and other performance measures.	In Fig. \ref{fig:JSI_Comp}, we compared the JSI scores produced by the proposed methods.
				\begin{table*}[]
					\centering
					\addtolength{\tabcolsep}{2pt}
					\renewcommand{\arraystretch}{1.5}
					\caption{Performance evaluation of different segmentation algorithms on PH2 dataset}
					\label{my-label3}
					\scalebox{0.95}{
						\begin{tabular}{cccccc}
							\hline
							User Name (Method)      & Accuracy & Dice &Jaccard Index &Sensitivity&Specificity  \\ \hline \hline
							FCN-16s   &0.917 &0.881    & 0.802 & 0.939 & 0.884 \\ 
							DeeplabV3+ &0.923 & 0.890    & 0.814  & 0.943 & 0.896\\  
							Mask-RCNN &0.937 & 0.904    & 0.830   & 0.969 & 0.897\\ 
							Ensemble-S (Proposed Method)   &\textbf{0.938} & \textbf{0.907}    & \textbf{0.839} & 0.932 & \textbf{0.929}  \\ 
							Ensemble-L (Proposed Method) &0.922 & 0.887    & 0.806 & 0.980 & 0.865  \\ 
							Ensemble-A (Proposed Method) &0.919 & 0.883    & 0.800 & \textbf{0.987} & 0.851  \\ \hline         
					\end{tabular}}
				\end{table*}

	\subsection{Comparison with the state of the art by lesion types}
	In ISIC-2017 segmentation task, the participants were asked to segment the boundaries of lesion irrespective of the lesion types. In this section, we compare the accuracy of segmentation results based on three lesion types: Naevus, Melanoma and Seborrhoeic Keratosis (SK).
		
	In Table \ref{tab:tradFeatsn} and \ref{tab:tradFeats}, we present the performance of our proposed method with other trained fully convolutional networks. We trained fully convolutional networks (FCN), DeepLabv3+, Mask-RCNN, and ensemble methods on the ISIC 2017 training set and tested on ISIC 2017 testing set. We observed that our proposed Ensemble-ADD method ourperformed in every category of the three lesion types.
	
	\subsection{Comparison on PH2 dataset}
	To test the robustness of our method and cross-dataset performance, we evaluate our proposed algorithms on PH2 dataset. It is worth noted that Ensemble-A produced the best results in ISIC 2017 testing set where as in PH2 dataset, Ensemble-S achieved better score in PH2 dataset, as shown in Table \ref{my-label3}.

	\section{Conclusion}
	Robust end-to-end skin segmentation solutions are very important to provide inference according to the ABCD rule system for the lesion diagnosis of melanoma. In this work, we proposed the fully automatic ensemble deep learning methods which combine one of the best segmentation methods, i.e. DeepLabv3+ (semantic segmentation) and Mask-RCNN (instance segmentation) to produce notably more accurate results than single-class segmentation CNN methods. We evaluated the performances on the ISIC 2017 testing set and PH2 dataset. We also utilized the pre-processing by using a colour constancy algorithm to normalize the data and then, morphological image functions for post-processing to produce segmentation results. Our proposed method outperformed the other state-of-the-art segmentation methods and 2017 ISIC challenge winners with good improvment on popular performance metrics used for segmentation. Further improvement can be made by fine-tuning the hyper-parameters of both networks in our ensemble methods. This study only focuses on the ensemble methods for segmentation tasks on skin lesion datasets. It can be further tested on the other publicly available segmentation datasets in both medical and non-medical domains. 
				
	% if have a single appendix:
	%\appendix[Proof of the Zonklar Equations]
	% or
	%\appendix  % for no appendix heading
	% do not use \section anymore after \appendix, only \section*
	% is possibly needed
	
	% use appendices with more than one appendix
	% then use \section to start each appendix
	% you must declare a \section before using any
	% \subsection or using \label (\appendices by itself
	% starts a section numbered zero.)
	%

	%\appendices
	%\section{Proof of the First Zonklar Equation}
	%Appendix one text goes here.
	
	% you can choose not to have a title for an appendix
	% if you want by leaving the argument blank
	%\section{}
	%Appendix two text goes here.

	% use section* for acknowledgment
	%\section*{Acknowledgment}

	%The authors would like to thank Gerard Pons, Joan Mart{\'i}, Robert Mart{\'i}, and UDIAT Diagnostic Centre of the Parc Taul{\'i} %Corporation, Sabadell (Spain) for providing Dataset B.

	% Can use something like this to put references on a page
	% by themselves when using endfloat and the captionsoff option.
	\ifCLASSOPTIONcaptionsoff
	\newpage
	\fi

	% trigger a \newpage just before the given reference
	% number - used to balance the columns on the last page
	% adjust value as needed - may need to be readjusted if
	% the document is modified later
	%\IEEEtriggeratref{8}
	% The "triggered" command can be changed if desired:
	%\IEEEtriggercmd{\enlargethispage{-5in}}
	
	% references section
	
	% can use a bibliography generated by BibTeX as a .bbl file
	% BibTeX documentation can be easily obtained at:
	% http://mirror.ctan.org/biblio/bibtex/contrib/doc/
	% The IEEEtran BibTeX style support page is at:
	% http://www.michaelshell.org/tex/ieeetran/bibtex/
	\bibliographystyle{IEEEtran}
	% argument is your BibTeX string definitions and bibliography database(s)
	\bibliography{skin}
\end{document}